\documentclass[runningheads]{svmult}

\usepackage{makeidx}   
\usepackage{graphicx}  
\usepackage{subeqnar}  
\usepackage{multicol}  
\usepackage{physprbb}  
\makeindex             

\begin{document}
\title*{Major Mergers and the Origin of Elliptical Galaxies}
\toctitle{Major Mergers and the Origin of Elliptical Galaxies}
\titlerunning{Elliptical Galaxies}
%
\author{Andreas Burkert\inst{1}
\and Thorsten Naab\inst{2}}
\authorrunning{Andreas Burkert and Thorsten Naab}
%
%
\institute{$^1$Max-Planck-Institut f\"ur Astronomie, K\"onigstuhl 17,\\
       D-69117 Heidelberg, Germany\\
$^2$Institute of Astronomy, Madingley Road, Cambridge CB3 0HA, UK}

\maketitle       

\begin{abstract}
The formation of elliptical galaxies as a result of the merging of spiral galaxies
is discussed. We analyse a large set of numerical N-Body merger simulations which 
show that major mergers can in principle explain the observed isophotal fine structure of 
ellipticals and its correlation with kinematical properties. Equal-mass mergers lead to
boxy, slowly rotating systems, unequal-mass mergers produce fast rotating and disky ellipticals.
However, several problems remain. 
Anisotropic equal mass mergers appear under certain projections disky  which
is not observed. The intrinsic ellipticities of remnants are often larger than observed. Finally, 
although unequal-mass mergers produce fast rotating ellipticals, the remnants are in general more anisotropic 
than expected from observations.
Additional processes seem to play an important role which are not included in dissipationless mergers.
They might provide interesting new information on the structure and gas content of
the progenitors of early-type galaxies.
\end{abstract}

\section{Introduction}

Giant elliptical galaxies are believed to be very old stellar systems that
formed by a major merger event preferentially
very early at a high redshift of more than two (Searle et al. 1973; Toomre \& Toomre 1972).
The merger triggered an intensive star-formation phase which
turned most of the gas of the progenitors into stars. Some fraction of the gas
was heated to temperatures of order the virial temperature, producing
X-ray coronae which are still visible today. The stellar disks of the progenitors were destroyed
as a result of the strong tidal forces during the merger, leading
to kinematically hot, spheroidal stellar remnants.
Subsequently, the systems
experienced very little accretion and merging with negligible star formation (Bruzual \& Charlot 1993). 
This scenario is supported by many observations which indicate that ellipticals contain
stellar populations that are compatible with purely passive evolution 
(Bower et al. 1992; Aragon-Salamanca et al. 1993; Ellis et al. 1997; Ziegler \& Bender 1997).
or with models of an exponentially, fast decreasing star formation rate (Ziegler et al. 1999).

An alternative scenario which is based on hierarchical theories of galaxy formation 
predicts that massive galaxies are assembled relatively late in many generations of
mergers through multiple mergers of small subunits, with additional smooth
accretion of gas (Kauffmann 1996; Kauffmann \& Charlot 1998). In this case, ellipticals might form either
if the multiple subunits are already preferentially stellar or if star
formation was very efficient during the protogalactic collapse phase (Larson 1974).

The idea that ellipticals form from major mergers of massive disk galaxies has been
originally proposed by Toomre \& Toomre (1972). Their ``merger hypothesis``  has been
explored in details by many authors, using numerical simulations.
Gerhard (1981), Negroponte \& White (1983), Barnes (1988) and Hernquist (1992) performed the
first fully self-consistent merger models of two equal-mass stellar
disks embedded in dark matter  halos. The remnants are slowly
rotating, pressure supported and anisotropic. They generally
follow an $r^{1/4}$ surface density profile for
radii $r \geq 0.5 r_e$, where $r_e$ is the effective radius. 
However it turns out that due to phase space limitations (Carlberg 1986),
an additional massive central bulge component is required
(Hernquist, 1993b), to fit the observed de Vaucouleurs profile (Burkert 1993) also in
the inner regions. All simulations demonstrated consistently that the global properties of
equal mass merger remnants resemble those of ordinary slowly rotating
massive elliptical galaxies. 

More recently it has become clear that ellipticals have quite a variety of fine structures with
peculiar kinematical properties which, in contrast to 
their universal global properties, can give a more detailed insight into their formation history.
It is interesting to investigate whether the merging hypothesis can explain these observations and,
if yes, whether they provide more information on the validity of this scenario,
the orbital parameters of the mergers and the 
structure and gas content of the progenitors from which the ellipticals formed.

Elliptical galaxies can be subdivided into two major groups with respect to their
structural properties (Bender et al. 1988; Bender 1988a,b; Kormendy \& Bender 1996).
Faint ellipticals are isotropic rotators with small minor
axis rotation and  disky deviations of their isophotal shapes from
perfect ellipses. Their isophotes are peaked in the rotational plane
and a Fourier analyses of the isophotal deviation from a perfect ellipse leads to a positive value of
the fourth order coefficient $a_4$. These galaxies might contain secondary, 
faint disk components which contribute up to 30\%
to the total light in the galaxy, indicating
disk-to-bulge ratios that overlap with those of S0-galaxies
(Rix \& White 1990; Scorza \& Bender 1995). Disky ellipticals have power-law inner density
profiles (Lauer et al. 1995; Faber et al. 1997) and show little or no radio and X-ray emission
(Bender et al. 1989). Most massive ellipticals have boxy isophotes, with negative values of $a_4$.
They also show flat cores (Lauer et al. 1995; Faber et al. 1997) and their kinematics is more complex
than that of disky ellipticals.  Boxy ellipticals rotate slowly, are
supported by velocity anisotropy and have a large amount of minor axis
rotation. Like the secondary disks of disky ellipticals, the boxy systems occasionally reveal
kinematically decoupled core components, that most likely formed from 
gas that dissipated its orbital energy during the merger, accumulated in the center and subsequently
turned into stars (Franx \& Illingworth 1988; Jedrzejewski \& Schechter 1988; Bender 1988a). 
The cores inhibit flattened rapidly   
rotating disk- or torus-like stellar structures that dominate the light in the
central few hundred parsecs (Rix \& White 1992, Mehlert et al.1998), but they
contribute only a few percent to the total light of the galaxy. The
fact that the stars are metal-enhanced confirms that gas infall and subsequently violent star formation,
coupled with metal-enrichment must have played an important role in the centers of
merger remnants (Bender \& Surma 1992; Davies et al. 1983; Bender et al. 1996; Davies 1996). 
Boxy ellipticals show strong
radio emission and high X-ray luminosities, resulting from emission from hot gaseous 
halos (Beuing et al. 1999) that probably formed from gas
heating during the merger. These hot gaseous bubbles are however absent in disky ellipticals. 
The distinct physical properties of disky and boxy elliptical galaxies
indicate that both types of ellipticals experienced different
formation histories. 

In order to understand the origin of boxy and disky ellipticals the isophotal
shapes of the numerical merger remnants have been investigated in detail. It has been 
shown that the same remnant can
appear either disky or boxy when viewed from different directions
(Hernquist 1993b) with a trend towards boxy isophotes (Heyl et al. 1994; Steinmetz \& Buchner 1995).
Barnes (1998) and Bendo \& Barnes (2000) analysed a 
sample of disk-disk mergers with a mass ratio of 3:1 and found that the
remnants are flattened and fast rotating in contrast to equal mass
mergers. Naab et al. (1999) studied the photometrical and kinematical
properties of a typical 1:1 and 3:1 merger remnant in details and
compared the results with observational data .
They found an excellent agreement and proposed that fast
rotating disky elliptical galaxies can originate from purely  
collisionless 3:1 mergers while slowly rotating, pressure supported
ellipticals form from equal mass mergers of disk galaxies. 

Despite these encouraging results no systematic high-resolution survey of mergers
has yet been performed to explore the parameter space of initial conditions and specify
the variety of properties of merger remnants that could arise.
Recently, Naab \& Burkert (2003) completed a large number of
112 merger simulations of disk galaxies adopting a statistically unbiased sample of
orbital initial conditions with mass ratios $\eta$ of 1:1,
2:1, 3:1, and 4:1. This large sample allows a much
more thorough investigation of the statistical properties of
merger remnants in comparison with observed disky and boxy ellipticals.

\section{The merger models}\label{mmod}
Cosmological simulations currently are not sophisticated enough to predict
initial conditions of major spiral mergers. Some insight can however
be gained by investigating the typical conditions under which dark matter halos merge
in standard cold dark matter models. Such a detailed analysis was done by
Khochfar, Burkert \& White (2003). The first encounter is in most cases a
parabolic orbit with an impact parameter of order the scale radius of the more massive dark halo, with 
random orientation of the net spin
axes of the progenitors. Unequal mass mergers with mass ratios $\eta$ of 3:1 to 4:1 are
as likely as equal-mass mergers with $\eta=1:1-2:1$. The cold dark matter simulations however do not provide
information on the internal structure and gas content of the merging spirals. In fact, simulations 
of hierarchical structure formation including
gas lead to disk galaxies which do not fit the zero point of the Tully-Fisher relation with
disk scale radii that are up to a factor of 10 smaller than observed (Navarro \& Steinmetz 2000).
Unless these problems are solved we cannot study the subsequent merging of disk galaxies
self-consistently, including the large-scale evolution of the Universe. In the meantime, the best
strategy is to construct plausible equilibrium models of 
disk galaxies and investigate their merging in isolation.

Equilibrium spirals were generated using the method described by Hernquist (1993a).
The following units are adopted: gravitational constant G=1, exponential
scale length of the larger disk $h=1$ and mass of the larger disk $M_d=1$.
For a typical spiral like the Milky Way these units correspond to $M_d=5.6 \times 10^{10} M_{\odot}$,
h=3.5 kpc and a unit time of $1.3 \times 10^7$ yrs.
Each galaxy consists of an exponential disk, a spherical, non-rotating
bulge with mass $M_b = 1/3$ and a Hernquist density profile (Hernquist 1990)
with a scale length $r_b=0.2$. The stellar system is embedded in a spherical
pseudo-isothermal halo with a mass $M_d=5.8$, cut-off radius $r_c=10$
and core radius $\gamma=1$. 

The mass ratios $\eta$ of the progenitor disks were varied between
$\eta =1$ and $\eta =4$. For equal-mass mergers ($\eta=1$)
in total 400000 particles were adopted with each galaxy
consisting of 20000 bulge particles, 60000 disk particles, and 120000
halo particles.  Twice as many halo particles 
than disk particles are necessary in order to reduce heating and instability effects in the
disk components (Naab et al. 1999). For the mergers with $\eta =2,3,4$
the parameters for the more massive galaxy were as described above. The
low-mass companion however contained a fraction of $1 / \eta $ less mass and
number of particles in each component, with a disk scale length of
$h=\sqrt{1/\eta }$, as expected from the Tully-Fisher relation (Pierce \& Tully 1992).

The N-body simulations for the equal-mass mergers were performed by
direct summation of the forces using the special purpose hardware
GRAPE6 (Makino et al. 2003).  The mergers with mass ratios $\eta =2,3,4$ were 
followed using the newly developed treecode WINE (Wetzstein et al.
2003) in combination with the GRAPE5 (Kawai et al. 2000) hardware.
WINE uses a binary tree in combination with the refined
multipole acceptance criterion proposed by Warren \& Salmon (1996). This criterion
enables the user 
to control the absolut force error which is introduced by the tree
construction. We chose a value of 0.001 which guarantees that the
error resulting from the tree is of order the intrinsic force error of
the GRAPE5 hardware which is $0.1\%$. 
For all simulations we used a gravitational softening of $\epsilon =
0.05$ and a fixed leap-frog integration time step of $\Delta t =
0.04$. For the equal-mass mergers simulated with direct summation on
GRAPE6 the total energy is conserved. The treecode in combination
with GRAPE5 conserves the total energy up to $0.5\%$. 

For all mergers, the galaxies approached each other on parabolic
orbits with an initial separation of $r_{sep} = 30$ length units and a pericenter    
distance of $r_p = 2$ length units. Free parameters are the inclinations of the
two disks relative to the orbital plane and the
arguments of pericenter. In selecting
unbiased initial parameters for the disk inclinations we followed the
procedure described by Barnes (1998). To determine the spin vector of each
disk we define four different orientations pointing to every vertex of
a regular tetrahedron.  These parameters result in 16 initial configurations
for equal mass mergers and 16 more for each mass ratio $\eta =2,3,4$
if the initial orientations are interchanged.  In total we simulated 112 mergers.      

In all simulations the merger remnants were allowed to settle into
equilibrium approximately  8 to 10 dynamical times after the
merger was complete. Then their equilibrium state was analysed.

\section{Photometric and kinematical properties of the remnants}
\label{pkp}
To compare our simulated merger remnants with observations we analysed
the remnants with respect to observed global photometric and
kinematical properties of giant elliptical galaxies, e.g. surface 
density profiles, isophotal deviation from perfect ellipses, velocity
dispersion, and major- and minor-axis rotation. Defining characteristic
values for each projected remnant we followed as closely as possible the
analysis described by Bender et al. (1988a).

\begin{figure}[h]
\begin{center}
\includegraphics[width=.8\textwidth]{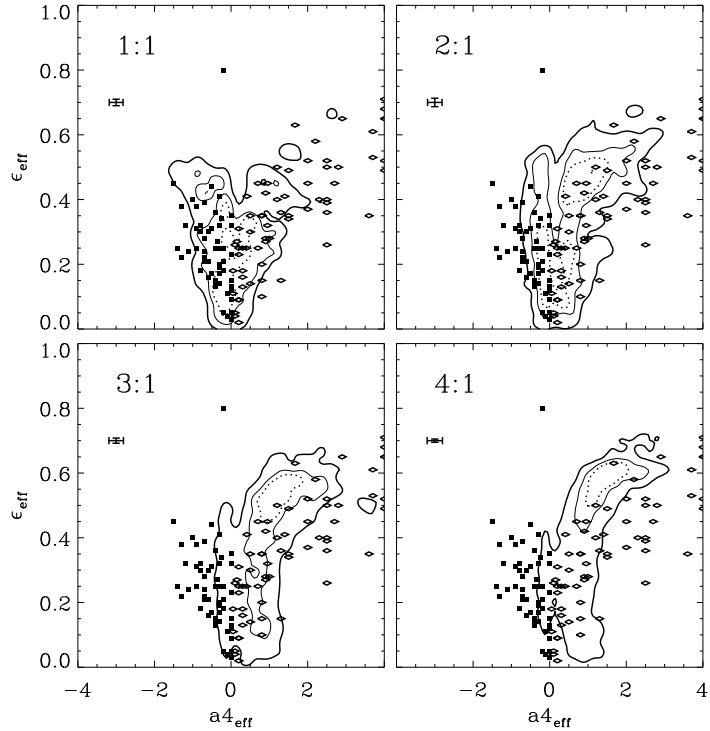}
\end{center}
\caption[]{Ellipticities versus fourth-order Fourier coefficient of the isophotal shape deviations is shown
for simulations with different initial mass ratios. The contours indicate the 50\% (dotted line), 70\%
(thin solid line) and the 90\% (thick solid line) probability to find a merger remnant in the enclosed area. Black squares indicate values
for observed boxy ellipticals, open diamonds show observed disky ellipticals.}
\label{fig1}
\end{figure}

\subsection{Isophotal shape}
An artificial image of the remnant was created by binning the central 
10 length units into $128 \times 128$ pixels. This picture was
smoothed with a Gaussian filter of standard deviation 1.5 pixels. The
isophotes and their deviations from perfect ellipses were then
determined using a data reduction package kindly provided by Ralf
Bender. Following the definition of Bender et al. (1988a) for the global properties of
observed giant elliptical galaxies, we determined for every
projection the effective $a_4$-coefficient $a4_{\mathbf{eff}}$ as the
mean value of $a_4$ between $0.25 r_e$ and $1.0 r_e$, 
with $r_e$ being the projected spherical half-light radius.  Like for observed ellipticals
we find two types of remnants. Disky systems show a positive characteristic peak of $a_4$
roughly at $0.5 r_e$. In boxy ellipticals, the $a_4$ coefficient might be positive in the innermost
regions. It decreases however systematically outwards with a mean value that is negative.
The characteristic ellipticity $\epsilon_{\mathbf{eff}}$ for each projection
was defined as the isophotal ellipticity at $1.5 r_e$.
To investigate projection effects we determined for each simulation
$a4_{\mathbf{eff}}$ and $\epsilon_{\mathbf{eff}}$ for 500 random
projections of the remnant. These values were used to calculate the two-dimensional
probability density function for a given simulated remnant to be "observed" in the
$a4_{\mathbf{eff}}$-$\epsilon_{\mathbf{eff}}$ plane. 

Figure 1 shows the ellipticities and $a_4$-coefficients of mergers with
$\eta = 1,2,3$, and $4$. The contours indicate the areas of 50\%
(dashed line), 70\% (thin line) and 90\% (thick line) probability to
detect a merger remnant with the given properties. Observed data
points from Bender et al. (1992) are over-plotted. Filled boxes are observed boxy ellipticals with
$a4_{\mathbf{eff}} \le 0$ while open diamonds indicate observed disky ellipticals with
$a4_{\mathbf{eff}} > 0$. The error bar in determining $a_4$ from the simulations
is shown in the upper left corner and was estimated applying the
statistical bootstrapping method  (Heyl et al. 1994).
Ellipticity errors are in general too small to be visible.

\begin{figure}[h]
\begin{center}
\includegraphics[width=.8\textwidth]{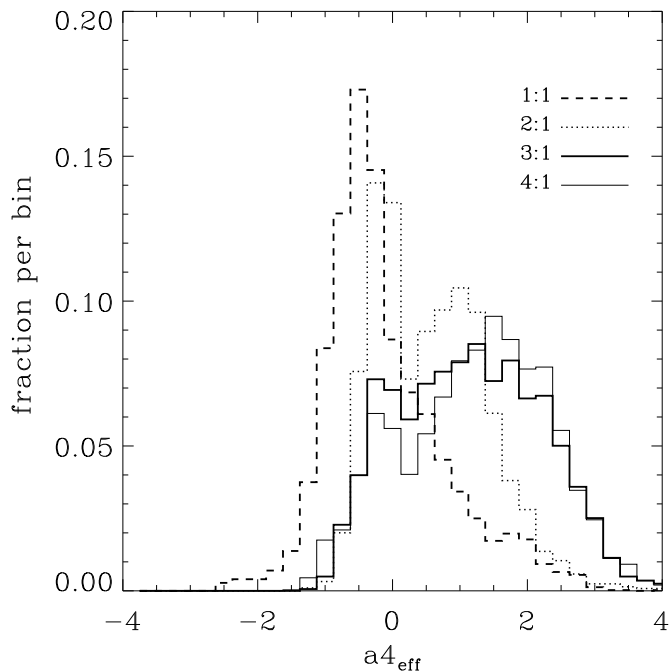}
\end{center}
\caption[]{Normalized histograms of the shape parameter $a4_{\mathbf{eff}}$ for mergers with various mass ratios.}
\label{fig2}
\end{figure}

We find that the isophotal shapes of ellipticals and their ellipticities are affected
by the initial mass ratio of the merger and by projection effects. 
The area covered by 1:1 remnants with negative
$a4_{\mathbf{eff}}$ is in very good 
agreement with the observed data for boxy elliptical galaxies. In
particular the observed trend for more boxy galaxies to have 
higher ellipticities is reproduced. However we also find configurations
of 1:1 mergers which under certain projection angles appear disky
with $0 \geq a4_{\mathbf{eff}} \leq 1$.
In addition, note that the remnants with $a4_{\mathbf{eff}}$
around zero can have higher ellipticities than observed. 

The distribution function of isophotal shapes for 1:1 merger remnants peaks at
$a4_{\mathbf{eff}} \approx -0.5\%$ (dashed curve in Fig. 2). It declines rapidly for more negative values
and has a broad wing towards positive $a4_{\mathbf{eff}}$ values. Almost half of the projected
remnants are disky. In contrast,
remnants of mergers with higher mass ratios shift in the direction of positive $a4_{\mathbf{eff}}$.
2:1 remnants peak at $a4_{\mathbf{eff}} \approx 0$. Now, 75\% of the projected remnants
show disky isophotes. For these cases, the observed trend of more disky ellipticals to be
more flattened is also clearly visible in Fig. 1. 3:1 and 4:1 mergers peak
at $a4_{\mathbf{eff}} \approx 1$. Their fraction of boxy projections is only 11\% and 7\%, respectively.
The very high positive values of $a4_{\mathbf{eff}} \geq 4\%$ observed in some ellipticals cannot be reproduced.
One might argue that these objects formed from mergers with even higher mass ratios of $\eta \geq 5:1$. 
However, in this
case, test simulations show that the merger remnants do not look like typical ellipticals anymore
with characteristic de Vaucouleurs profiles 
as the more massive disk is not destroyed. Their surface brightness profiles instead
remain exponential.

In summary, there is a clear trend for unequal-mass mergers to 
produce more disky remnants. Responsible for the disky appearance of
the 3:1 and 4:1 remnants is the distribution of the particles of the
massive disk (Barnes 1998).  The particles originating from the
small progenitor accumulate in a torus-like structure with
peanut-shaped or boxy isophotes while the luminous material of the
larger progenitor still keeps its disk-like appearance. In combination,
the contribution from the larger progenitor -- since it is three to four times
more massive -- dominates the overall appearance of the remnant. This
result holds for all 3:1 and 4:1 merger remnants. 
For equal mass
mergers however both disks are destroyed efficiently during the merger. No dominant disk-like structure
remains after the merger and the system looses the information about the initial configuration.

\begin{figure}[h]
\begin{center}
\includegraphics[width=.8\textwidth]{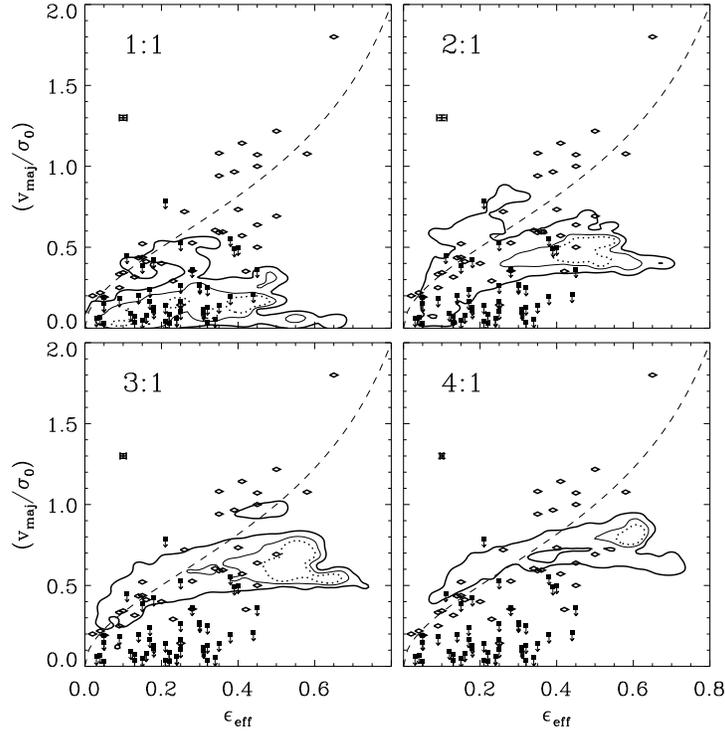}
\end{center}
\caption[]{Rotational velocity over velocity dispersion versus characteristic ellipticity for mergers with various
mass ratios. Values for observed ellipticals are overplotted. The dashed line shows the theoretically
predicted correlation for an oblate isotropic rotator.}
\label{fig3}
\end{figure}

\subsection{Kinematics}

The central velocity dispersion $\sigma_0$ of every remnant
is determined as the average projected velocity dispersion of the
stars inside a projected 
galactocentric radius of $ 0.2 r_e $. The
characteristic rotational velocity $v_{maj}$ along the major axis is defined
as the projected rotational velocity determined around $1.5 r_e$.
Like for the isophotal shape we constructed probability density plots
for the kinematical properties of the simulated remnants and compared
them with observational data from elliptical galaxies. Figure 3 shows the
distribution function in the
$(v_{maj}/\sigma_0)$-$\epsilon_{\mathbf{eff}}$ plane.

\begin{figure}[h]
\begin{center}
\includegraphics[width=.8\textwidth]{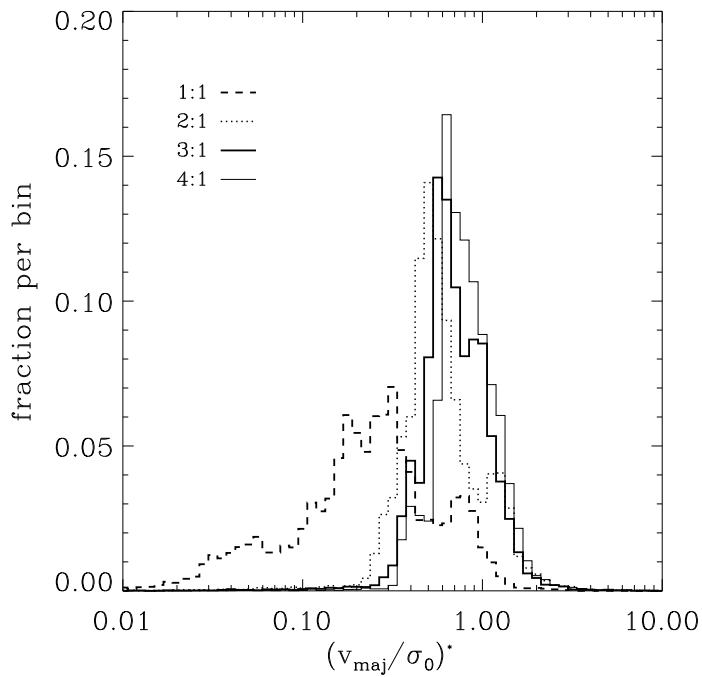}
\end{center}
\caption[]{Normalized histograms of $(v_{maj}/\sigma_0)^*$ for 1:1 (dashed line), 2:1 (dotted line), 3:1 (thick line)
and 4:1 (thin line) mergers.}
\label{fig4}
\end{figure}

The region of slowly rotating boxy ellipticals (filled squares) is almost completely
covered by the data of 1:1 mergers. Unequal-mass merger remnants are clearly fast rotating.
They can be associated with disky
ellipticals.  Although the simulated remnants are in good agreement with
observations there is again the trend for the 
ellipticities to be higher than observed, especially when the system is seen edge-on.

\begin{figure}[h]
\begin{center}
\includegraphics[width=.8\textwidth]{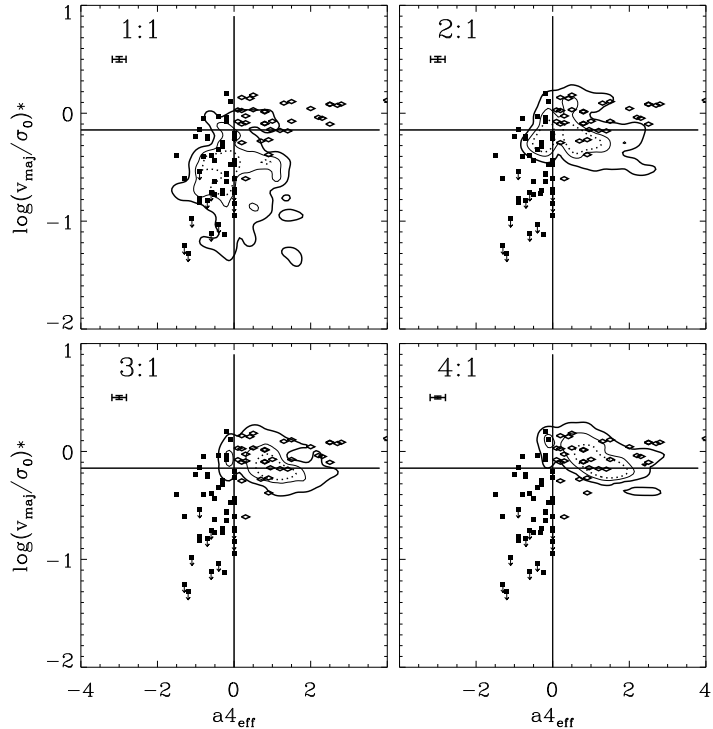}
\end{center}
\caption[]{Anisotropy parameter versus isophotal shape for mergers with various mass ratios.
Values for observed ellipticals are overplotted.}
\label{fig5}
\end{figure}

The anisotropy parameter  $(v_{maj}/\sigma_0)^*$ is defined as the
ratio of the observed value of  $(v_{maj}/\sigma_0)$
and the theoretical value for an isotropic oblate rotator
$(v/\sigma)_{theo} = \sqrt{\epsilon_{\mathbf{obs}} / 
(1-\epsilon_{\mathbf{obs}})}$ with the observed
ellipticity $\epsilon_{\mathbf{obs}}$ (Binney 1978). This parameter is frequently used
by observers to test whether a given galaxy is flattened by rotation
($(v_{maj}/\sigma_0)^* \ge 0.7$)  or by velocity anisotropy
($(v_{maj}/\sigma_0)^* < 0.7$) (Davies et al. 1983; Bender 1988a; Nieto et al. 1988;
Scorza \& Bender 1995).  Figure 4 shows the 
normalized  histograms for the $(v_{maj}/\sigma_0)^*$ values of the simulated
remnants. 1:1 remnants peak around
$(v_{maj}/\sigma_0)^* \approx 0.3$ with a more prominent tail towards
lower values.  They are consistent with
being supported by anisotropic velocity dispersions. As these systems also have
preferentially negative $a_4$-values they agree with observations of boxy ellipticals (Fig. 5).
Unequal mass mergers peak at $(v_{maj}/\sigma_0)^*
\approx 0.7$, as expected for oblate isotropic rotators. 
Since especially the 3:1 and 4:1 remnants also have predominantly disky isophotes they cover the area   
populated by observed disky ellipticals in the 
$log (v_{maj}/\sigma_0)^*$ - $a4_{\mathbf{eff}}$ diagram which is shown in figure 5.

We also investigated the minor-axis kinematics of the simulated remnants by determining the rotation
velocity along the minor axis at $0.5 r_{eff}$. The amount of minor axis rotation 
was characterized by $(v_{min} /
\sqrt{v_{maj}^2 + v_{min}^2})$ (Binney 1985). Minor axis rotation in
elliptical galaxies, in addition to isophotal twist, has been
suggested as a sign for a triaxial shape of the main body of
elliptical galaxies (Wagner et al. 1988; Franx et al. 1991). In general,
1:1 mergers show a significant amount of minor-axis rotation, whereas
3:1 and 4:1 remnants have only small minor axis rotation (for details see Naab \& Burkert 2003).

\section{Conclusions}
\label{con}
The analysis of a large set of mergers with different mass ratios and orbital
geometries shows that their properties are in general in good agreement with the observational data 
for elliptical galaxies. 

Only equal mass mergers can produce boxy, anisotropic and slowly 
rotating remnants with a large amount of minor axis rotation. However, in the more
unlikely case that the initial spins of the progenitor disks are aligned, the remnants appear 
isotropic and disky or boxy depending on the orientation.  
In contrast, 3:1 and 4:1 mergers form a more homogeneous group of remnants. 
They have preferentially disky isophotes, are always fast rotating and show 
small minor axis rotation independent of the assumed projection. 
2:1 mergers have properties intermediate between
boxy or disky ellipticals, depending on the projection and the
orbital geometry of the merger. 

There still exist problems which are not solved up to now.
Certain projections of 1:1 mergers lead to 
anisotropic, disky remnants which are not observed.
Edge-on projections of merger remnants often show very high 
ellipticities $\epsilon > 0.6$ which are larger than observed. Finally, 
some 2:1 to 4:1 remnants are more anisotropic than expected from their rotation.
Their values of $(v/\sigma)^*$ are smaller and their ellipticities are larger 
than observed. A problem arises especially for very
low luminosity giant ellipticals which are characterized by exceptionally high rotational velocities in the outer regions
that cannot be reproduced (Cretton et al. 2001).
A detailed analyses of the intrinsic kinematics of disky, fast rotating merger unequal-mass remnants which are
called isotropic due to their high $(v/\sigma)^* \approx 0.7$
demonstrates that in most cases the velocity dispersion tensor is as anisotropic as for equal-mass, boxy and
anisotropic mergers with $(v/\sigma)^* =0.1$ (Burkert et al. 2003). The anisotropy parameter therefore is
not necessarily a good indicator of anisotropy. It rather measures the amount of rotation in the systems.

The present simulations were purely dissipationless, taking into
account only the stellar and dark matter components. The importance of gas
in determining the structure of merger remnants is
not clear up to now.  Kormendy \& Bender (1996) proposed a revised Hubble
sequence with disky ellipticals representing the missing link between
late type systems and boxy ellipticals. They noted that
gas infall into the equatorial plane with subsequent star formation
could explain the origin of diskyness. Scorza \& Bender (1995)
 demonstrated that ellipticals with embedded disks
would indeed appear disky when seen edge-on and boxy
otherwise.  Although this scenario appears attractive, it cannot explain why X-ray halos are
found only in boxy ellipticals. As the detection of hot gas around galaxies should be 
independent of their orientation, the isophotal shapes
of ellipticals would not correlate with their X-ray emission, if these shapes
are merely a result of projection effects.

Our simulations indicate that it is preferentially the initial mass ratio which
determines the isophotal shapes of merger remnants. Still, gas could have played an important role
in affecting the final structure and stellar population of ellipticals (Bekki 1998, 1999; Mihos \& Hernquist 1996), not only in their
central regions and might solve the problem of dissipationless mergers.  Naab \& Burkert (2001a) have shown that
extended gas disks can form as a result of a gas rich unequal mass
mergers (see also Barnes 2002).  Naab \& Burkert (2001b) investigated
line-of-sight velocity distributions of dissipationless merger
remnants and found a velocity profile asymmetry that is opposite to
the observations. They concluded that this disagreement can be solved
if ellipticals would indeed contain a second disk-like substructure that most
likely formed through gas accretion. The situation is
however not completely clear, as another study by Bendo \& Barnes (2000) found
a good agreement of the observed asymmetries for some cases. More
simulations, including gas and star formation will be required to
understand the role of gas in mergers and to answer the question of how early-type galaxies formed.

{\bf Acknowledgement:} A. Burkert thanks the organizers of the workshop
"Galaxies and Chaos" for the invitation and a very stimulating conference.

\end{document}